\documentclass[
singlecolumn,nofootinbib,
 amsmath,amssymb,
 aps,
]{revtex4-2}
\usepackage{graphicx}% Include figure files
\usepackage{dcolumn}% Align table columns on decimal point
\usepackage{bm}% bold math
\usepackage{caption}%for subfigures
\usepackage{subcaption}%for subfigures
\usepackage{xcolor}%for colored text(comments)
\usepackage{color,soul}%for colored text(comments)

\usepackage[normalem]{ulem}

\setstcolor{red}

\begin{document}

\preprint{APS/123-QED}

\title{
The $L_\mu-L_\tau$ solution to the IceCube Ultra High Energy neutrino deficit in light of NA64 }
%authors
\author{Leon M.G. de la Vega }%
 \email{leonmgarcia@phys.ncts.ntu.edu.tw}
\affiliation{%
Physics Division, National Center for Theoretical Sciences, National Taiwan University, Taipei 106319, Taiwan\\
 Instituto de F\'{\i}sica, Universidad Nacional Aut\'onoma de M\'exico, A.P. 20-364, Ciudad de M\'exico 01000, M\'exico.
 }%
\author{Eduardo Peinado}%
 \email{epeinado@fisica.unam.mx}
\affiliation{%
 Instituto de F\'{\i}sica, Universidad Nacional Aut\'onoma de M\'exico, A.P. 20-364, Ciudad de M\'exico 01000, M\'exico. }\affiliation{Departamento de F\'isica, Centro de Investigaci\'on y de Estudios Avanzados del Instituto Polit\'ecnico Nacional.}%
\author{Jos\'e Wudka }%
 \email{jose.wudka@ucr.edu}
\affiliation{%
 Department of Physics and Astronomy, UC Riverside, Riverside, California 92521-0413, USA.}%
\date{\today}% It is always \today, today,

\begin{abstract}
In this work we analyze the scenario where a MeV scale $L_\mu-L_\tau$ gauge boson can explain the deficit in the diffuse ultra high energy (UHE) astrophysical neutrino spectrum observed in IceCube, as well as the discrepancy between experimental and $e^+ e^-$ scattering data driven SM calculations of the muon anomalous magnetic moment. We map the parameter space of the model where the elastic resonant s-channel scattering of UHE neutrinos with the cosmic neutrino background, mediated by the new $Z^\prime$, can improve the description of the observed cascade and track spectra over the no-scattering hypothesis. Comparing to recent NA64-$\mu$ results, we find that some part of the parameter space remains unexplored, but at a data volume of $10^{11}$ muons on target NA64-$\mu$ will completely probe this region. 
\end{abstract}

\maketitle

%\tableofcontents

\section{Introduction}
Recently, the NA64 collaboration at LHC has released their first results on the search for light leptophilic gauge bosons in $\mu+N\rightarrow\mu+N +(Z^\prime\rightarrow \text{inv.})$  \cite{Andreev:2024sgn} and $e+N\rightarrow e+N +(Z^\prime\rightarrow \text{inv.})$ \cite{Andreev:2024lps} bremsstrahlung processes. NA64 has begun probing the parameter space where a light $L_\mu-L_\tau$ gauge boson (denoted by $Z'$) can account for the muon $(\text{g-2})_\mu$ magnetic  anomaly \cite{Andreev:2024lps,Andreev:2024sgn}. NA64 has the future potential to cover the whole region of parameter space that resolves the  $(\text{g-2})_\mu$ tension in these models,
with an estimated sensitivity to the $L_\mu-L_\tau$ gauge coupling of $\sim 4\times 10^{-5}$ for a $Z^\prime$ mass $1-100$ MeV, for a total run of $10^{12}$ muons on target (MOT) \cite{Sieber:2021fue}. 

It has been pointed out that a $L_\mu-L_\tau$ gauge boson in this mass range can have a noticeable impact on the diffuse astrophysical neutrino background as measured by IceCube \cite{Araki:2014ona,Kamada:2015era,DiFranzo:2015qea,Carpio:2021jhu,Hooper:2023fqn}, analogous to the expected effect of the SM Z boson resonance \cite{Weiler:1982qy}. As the effect relies on the resonant scattering of astrophysical neutrinos on cosmic neutrinos, it can be sizeable even for small values of the gauge coupling \cite{Ioka:2014kca,Cherry:2014xra,Creque-Sarbinowski:2020qhz}. 
Recently, references \cite{Carpio:2021jhu,Hooper:2023fqn} highlighted that the apparent deficit of neutrinos of energies $ \sim 10^6\,$GeV in the diffuse spectrum as measured by IceCube \cite{IceCube:2020acn} could be explained by the $L_\mu-L_\tau$ gauge boson that also drives the value of $(\text{g-2})_\mu$ away from the expected SM value,  obtained using electron scattering data for the hadronic vacuum polarization contribution when the $Z^\prime$ boson has a mass of order $\mathcal{O}(10 \text{MeV})$.  Motivated by the NA64 recent results and its future prospects, in this work, we explore the impact of this experiment on the $L_\mu-L_\tau$ solution to $(\text{g-2})_\mu$ and the above-mentioned neutrino deficit. We find that NA64 will not only completely probe the region where the muon g-2 anomaly is explained by a light $L_\mu-L_\tau$ $Z^\prime$, but also the region where the $Z^\prime$ explains the apparent IceCube neutrino deficit. 

\section{Neutrino Self-interactions in the diffuse neutrino astrophysical spectrum}
During the past decade, IceCube has measured an excess of the neutrino flux over the expected atmospheric neutrino spectrum at high energies with high statistical significance \cite{IceCube:2013low,IceCube:2014stg,IceCube:2016umi,IceCube:2015gsk}. The origin of this neutrino flux is still not well understood. Diverse astrophysical objects such as Active Galactic Nuclei have been proposed as sources of UHE neutrinos \cite{Murase:2022feu,Eichler:1979yy,Stecker:1991vm,Mannheim:1993jg}.
%; this excess is attributed to the emission of neutrinos by astrophysical objects, specifically Active Galactic Nuclei (AGN). 
The flux of neutrinos produced by AGNs is associated with hadronic production, acceleration, and subsequent decay. In these production mechanisms, muon and electron neutrinos are produced from pion decays. As the propagation and eventual detection timescale is much larger than the oscillation length, the neutrino flavor composition is expected to be equipartitioned among the three SM flavors.
The IceCube detector is sensitive to the three known flavors of neutrinos: muon neutrinos are primarily detected by the production of muon tracks passing through the detector, or originating within it. Electron and tau neutrinos produce cascade events with low atmospheric background and good energy resolution ($\sim 15 \%$). The IceCube collaboration has utilized data from these two types of events to report the observed energy spectrum of the astrophysical muon neutrino flux  \cite{IceCube:2021uhz} and the combined electron and tau neutrino flux \cite{IceCube:2020acn}. Furthermore, the IceCube collaboration has observed a Glashow resonance event \cite{IceCube:2021rpz}, i.e., $\bar{\nu}_e + e^-\rightarrow W^-\rightarrow \textit{anything} $, near a neutrino energy of $6.3$ PeV in the electron rest frame, where the $W$ resonance enhances the cross-section.
Interestingly, the $\nu_e + \nu_\tau$ spectrum observed by IceCube contains a hint of a deficit in the $200\text{ TeV} - 1 \text{ PeV}$ energy range, while the $\nu_\mu$ spectrum with its poorer energy resolution does not seem to contain such a deficit.  The observed dip in neutrino flux has been associated to resonant neutrino self-interactions mediated by hypothetical new light bosons \cite{Hyde:2023eph,Barenboim:2019tux,Chauhan:2018dkd,Kamada:2015era,Bustamante:2020mep}. 
In this work, we consider the possibility that such a deficit is the result of a resonant effect of the astrophysical neutrino -- cosmic neutrino scattering cross-section, induced by the presence of a hypothetical light $Z'$ gauge boson that couples to the (left-handed) SM neutrinos:
\begin{equation}
\mathcal{L}_{Z'}=\sum_{i=e,\mu,\tau} Q_i' Z'_\mu \bar{\nu}_i \gamma^\mu P_L\nu_i\; .
    \label{eq:couplings}
\end{equation}
For a $Z'$ mediator that couples to neutrinos, the scattering cross section for the $\nu_i \bar{\nu}_j \rightarrow \nu \bar{\nu}$ process is 
\begin{equation}
    \sigma_{ij} = \frac{2}{3\pi} g^4_{\mu\tau}Q_{ij}^2 \frac{s}{(s-m_{Z^\prime}^2)^2+\Gamma_{Z^\prime}^2 m_{Z^\prime}^2 } \quad,
\end{equation}
where $Q_{ij}=U_{\mu i}U_{\mu j} - U_{\tau i}U_{\tau j}  $. 
The decay width of the $Z'$, assuming $m_{Z^\prime}<2m_\mu$ and that the $Z^\prime$ couplings to electrons can be neglected, is 
\begin{equation}
\Gamma_{Z^\prime}=\frac{g_{\mu\tau}^4m_{Z^\prime}}{12\pi}
\end{equation}
The condition for the s-channel resonance is:
\begin{equation}
    E_\nu=\frac{ m_{Z'}^2}{2m_\nu} \; ,
    \label{eq:res.en}
\end{equation}
where $E_\nu$ is the energy of the astrophysical neutrino, $m_{Z'}$ is the $Z'$ gauge boson mass, and $m_\nu$ is the mass of the target, the cosmic neutrino background (C$\nu$B), (neglecting its velocity). Due to the $Z'$-mediated resonance, we expect a sharp reduction in the cosmic neutrino flux at the resonant energy Eq. (\ref{eq:res.en}). The process $\nu \bar{\nu}\rightarrow Z' Z'$ can contribute to the depletion of astrophysical neutrinos, but as this is a non-resonant process, it can only contribute significantly for gauge couplings of order $\mathcal{O}(1)$ \cite{Esteban:2021tub}.

The evolution of neutrino flux from production at sources  through propagation in the C$\nu$B until it reaches Earth is governed by a Boltzmann equation \cite{Ng:2014pca,Blum:2014ewa,Araki:2015dia,Creque-Sarbinowski:2020qhz} that determines the time dependence of the differential comoving neutrino density of a neutrino mass eigenstate $m_i$,  $\frac{d n_i(t)}{dE_\nu}=\hat{n}_i(t,E_\nu)$:  
\begin{equation}
\label{neutdens}
     \frac{\partial\hat{n}_i}{\partial t} =\frac{\partial}{\partial E_\nu} (H E_\nu \hat{n}_i ) +  \mathcal{L}_i(E_\nu,t) - \hat{n}_i\sum_j\sigma_{ij} n_j +\sum_{j,k,l} n_j\int_{E_\nu}^\infty d E_\nu^\prime\,   \frac{d\sigma_{jk\rightarrow il}(E_\nu,E_\nu^\prime)}{dE_\nu} \hat{n}_k(E_\nu',t)\quad ,
\end{equation}
where $H(t)$ is the Hubble rate, $\mathcal{L}_i$ is the neutrino luminosity (for the $i-$th flavor) of emitting bodies, $n_j(t)$ is the density of the C$\nu$B, and $d\sigma_{jk\rightarrow il}$ is the neutrino self scattering differential cross section. 

The first term on the right-hand side of eq. (\ref{neutdens}) accounts for the dilution of neutrino density due to the expansion of the Universe. The second term describes the appearance of neutrinos as sources emit them. The third and fourth terms describe the effect of neutrino self-scatterings in the spectrum. The third term removes the neutrinos with energy $E_\nu$ from the high energy spectrum as they scatter with the C$\nu$B. The fourth term is a regeneration term, which adds to the spectrum of the scattered neutrinos according to their acquired (or lost) energy.

\section{Limits on leptophilic $Z'$ bosons}

A $Z'$ gauge boson coupling to neutrinos as described in eq. (\ref{eq:couplings}) is subject to rigorous theoretical and experimental constraints. From the theoretical side, due to the $SU(2)_L$ symmetry of the SM neutrinos lie within a doublet with the (left-handed) charged leptons and thus the $\nu$-$Z'$ interactions of eq. (\ref{eq:couplings}) must be accompanied by the charged lepton couplings to the $Z'$. Charged lepton interactions with a light mediator are experimentally strongly constrained, especially through electron-electron and electron-neutrino scatterings.
%One might expect the left-handed charged-lepton and left-handed neutrino couplings to the $Z'$ to be of equal magnitude, but this need not be the case: a UV-mechanism suppressing charged-lepton couplings can be constructed, using the fact that the $SU(2)_L$ symmetry is spontaneously broken by the SM scalar sector. For example, in the presence of a gauge symmetry that acts on SM singlet fermions, which in turn mix with SM left-handed neutrinos in an Inverse seesaw scenario, one can obtain light neutrino couplings to the new gauge boson governed by active-sterile neutrino mixing, while the leading contributions to charged fermion couplings with the new gauge boson arise from gauge boson mixing and one loop effects \cite{Abdallah:2021npg,delaVega:2022uko}.

A gauge symmetry must satisfy the gauge anomaly cancellation conditions arising from triangle diagrams. This imposes conditions on the fermion representations under the gauge symmetry. For our analysis, we will consider a model where $L_\mu - L_\tau $ is gauged (we denote by $L_f,\, f=e,\,\mu,\,\tau$ the lepton number for each flavor), which is trivially anomaly-free \cite{He:1990pn,He:1991qd}. We select this model because it naturally avoids the constraints on possible $Z'$ couplings to first-generation leptons and because the $Z'$ coupling to the second-generation leptons may alleviate tensions in the measurement of the anomalous magnetic moment ($g-2$) for the muon. In this model, the strongest constraints when the $Z'$ mass is in the  $1-100$ MeV  range are:
\begin{enumerate}
    \item {\em Cosmology \& Astrophysics}: A light vector boson interacting with leptons can be produced in the early Universe; its subsequent decay would inject energy into the neutrino bath, delaying neutrino decoupling, thus affecting the observed value of BSM relativistic degrees of freedom $\Delta N_{\tt eff}$ \cite{Huang:2017egl,Escudero:2019gzq}. This constraint favors values of $M_{Z'}>9$ MeV for $g'>10^{-9}$, but it's also sensitive to the kinetic mixing between the new gauge boson and the photon \cite{Escudero:2019gzq}. A ``natural" value of the kinetic mixing, obtained from the renormalization group equations of the minimal $SM\times U(1)_{L_\mu-L_\tau}$ model allows  $M_{Z'}>9$ MeV. We also note that the extra contribution to $\Delta N_{\tt eff}$ can help alleviate the Hubble constant tension \cite{Escudero:2019gzq,Carpio:2021jhu}. 

    Recently, strong constraints on the $L_\mu-L_\tau$ gauge coupling for a sub-GeV $Z^\prime$ have been derived from limits on the cooling rate of observed white dwarfs \cite{Foldenauer:2024cdp}. Current limits disfavor the $Z^\prime$ as a solution to the muon g-2 anomaly for a $m_{Z^\prime}<9$ MeV, when the value of the kinetic mixing of the new gauge boson to the photon at low energies is taken to be \textit{natural}, $\epsilon\sim -g^\prime/70$ .
    \item {\em $Z,W$ decays}: The precise measurement of the $Z$ and $W$ decay widths with final state neutrinos can be used to set a limit on the partial width for a decay channel of the form $Z(W)\rightarrow \nu(\mu) Z^\prime \nu $\cite{Laha:2013xua}, that arise when a $Z^\prime$ lighter than the SM gauge bosons couples to the active neutrinos. The resulting $Z$ decay limit is applicable to any model with neutrinos coupling to the $Z^\prime$. The limit derived from the $W$ decay is stronger than that of the $Z$ but applies only when the $Z^\prime$ couples to the muon.  
    \item {\em Neutrino-electron / nucleus scattering}: One of the strongest constraints on new gauge bosons coupling to neutrinos come from limits on neutrino scattering cross sections. In the $L_\mu-L_\tau$ scenario, with natural values for the kinetic mixing \cite{Bauer:2022nwt}, strong constraints on the gauge coupling can be derived from the measurement of solar neutrino scattering on electrons at Borexino, and the scattering of neutrinos on nuclei on dark matter detectors \cite{Amaral:2020tga}. Future dark matter experiments such as DARWIN will also have the potential of completely probing the region where the $Z^\prime$ can explain the $(g-2)_\mu$ anomaly for a sub-60 MeV mass $Z^\prime$.
    \item {\em Kaon decays}: The strongest limit on gauged $L_\mu-L_\tau$ model with $ m_{Z'}$ in the sub-MeV region comes from Kaon decays \cite{Laha:2013xua}. The charged $K^+$ decays dominantly through a $W$ exchange process into a $\mu^+ \nu$ two-body state. The kinematics of the two body decay (combined with the initial $K^+$ momentum distribution) tightly constrain the emitted muon to a range of energies. In a model with a light $Z^\prime$ coupling to the muon and to the muon neutrino, the three-body decay $K^+\rightarrow Z^\prime \mu^+\nu$ is kinematically allowed. In this process the outgoing muon has an energy above the above range. The absence of muons produced in kaon decays with anomalous energies provides limits on the gauge coupling and mass of the $Z^\prime$ boson.
    \item \textit{Beam dump}: Searches for $L_\mu-L_\tau$ $Z^\prime$ production in NA64 strongly constrain the $1$ MeV to $100$ MeV region \cite{Andreev:2024lps,Andreev:2024sgn}. This experiment consists of an electron or muon beam dump on a fixed target. NA64 searches for missing energy signatures of the process $e+N \rightarrow e+N+(Z\rightarrow \text{inv.})$ in the electron beam mode (NA64$e$) and $\mu+N \rightarrow \mu+N+(Z\rightarrow \text{inv.})$ in the muon beam mode (NA64$\mu$). The results for the muon beam mode with $1.98 \times 10^{10}$ MOT already start to probe the region of the parameter space where the $Z^\prime$ can explain the muon $g-2$ anomaly, disfavoring a mass $m_{Z^\prime}$ larger than $30$ MeV as a solution to the anomaly.
\end{enumerate}

\subsection{Muon anomalous magnetic moment}
The experimental value of the anomalous magnetic moment of the muon ($a_\mu$) as measured by the Muon g-2 experiment at Fermilab \cite{Muong-2:2021ojo} is in tension with the SM theoretical prediction as calculated by the Muon g-2 Theory Initiative White Paper \cite{Aoyama:2020ynm}. The experimental value reports an excess of 4.2 $\sigma$ over the White Paper (WP) prediction: 
\begin{equation}
    a_\mu^{\text{exp}} -a_\mu^{\text{SM(WP)}} = (25.1 \pm 5.9)\times 10^{-10} \quad .
\end{equation}
A significant source of error in the WP estimate lies in the calculation of the Hadronic Vacuum Polarization (HVP), which is obtained through a data-driven method using the measured cross sections of $e^+e^- \rightarrow \text{hadrons}$ processes. An alternate method of obtaining the HVP contribution to $a_\mu$ is through Lattice QCD methods. Several collaborations have published calculations with error estimates comparable to the data-driven method \cite{Borsanyi:2020mff,FermilabLatticeHPQCD:2023jof,RBC:2023pvn}, yielding a larger value of the HVP contribution, thereby reducing the tension between theoretical prediction and experimental measurement to $\sim 1.5\sigma$\footnote{Additionally, recent measurements of the $e^+e^- \rightarrow \text{hadrons}$ cross sections by the CMD-3 collaboration at the VEPP-2000 collider have resulted in a larger theoretical value of the HVP contribution to $a_\mu$, reducing the theoretical discrepancy compared to the WP result \cite{CMD-3:2023alj}. }. The g-2 tension becomes
\begin{equation}
     a_\mu^{\text{exp}} -a_\mu^{\text{SM(BMW)}} 
= (10.7 \pm 7.0)\times 10^{-10} \quad ,
\end{equation}
when the BMW collaboration results \cite{Borsanyi:2020mff} are used in place of the data-driven results for the HVP contribution in the WP estimate. The $L_\mu-L_\tau$ model can eliminate the muon g-2 WP tension with the $Z'$ one-loop contribution to the muon magnetic moment. The $Z'$ loop contribution to $a_\mu$ is given by
\begin{equation}
\Delta a_\mu^{Z'}=\frac{g'^2 m_\mu^2  }{4 \pi^2 m_{Z'}^2}\int_0^1 dx \frac{x^2(1-x)}{1-x+x^2\frac{m_\mu^2}{m_{Z'}^2}}
\end{equation}
To avoid experimental limits from direct $Z'$ searches, the sub-100 MeV $Z'$ mass is the only allowed region accommodating the WP result \cite{AtzoriCorona:2022moj}. On the other hand, if the HVP Lattice calculations are taken to be more accurate than the WP data-driven approach, the Muon g-2 results can be used to exclude a large part of the $L_\mu-L_\tau$ parameter space, potentially including parts of the region where the IceCube dips are explained by resonant neutrino interactions.
\section{Methodology}
We use the software nuSIprop \cite{Esteban:2021tub}, modified to include the vector boson interactions described in the preceding sections, to calculate the astrophysical neutrino spectrum after it propagates through the C$\nu$B. We parametrize the initial spectrum, assumed equal for all neutrino mass eigenstates, with the power law
\begin{equation}
   \frac{ \Phi_{\nu}}{C}=A \left(\frac{E_\nu}{100 \text{TeV}} \right)^{-\gamma} \; ,
\end{equation}
where $C=3\times 10^{-18} \text{GeV}^{-1}\text{cm}^{-2}\text{s}^{-1}\text{sr}^{-1}$; $A,\gamma$ are dimensionless parameters ($\gamma$ is commonly called the spectral index). This type of spectrum is expected for neutrino production in AGN from hadronic processes, themselves subject to power laws. We assume that AGNs are distributed in redshift following the star formation rate. The relevant parameters for our study are then ${m_{Z'},g',A,\gamma}$, the first two are the parameters of the gauge extension of the SM, while the last two are the astrophysical parameters of neutrino emission. We scan these parameters in the ranges
\begin{eqnarray}
    m_{Z'}&\in& [1,20] \text{ MeV} \\ \nonumber
    g'&\in& [10^{-5},10^{-1}] \\ \nonumber
    A &\in& [0.1,2.5] \\ \nonumber
\gamma&\in& [1.8,3.5]\quad ,
\end{eqnarray}
with the sum of neutrino masses fixed at $0.1 \text{ eV}$
and analyze the results using a chi-squared distribution. We follow the analysis method of \cite{Hooper:2023fqn}, defining 
\begin{equation}
\chi^2(m_{Z'},g',A,\gamma)=\sum_{i,\alpha}\frac{(N_{\text{th}}(i,\alpha)-N_{\text{obs}}(i,\alpha))^2}{(\sigma^i_{\text{obs}}(i,\alpha) )^2} \quad ,
\end{equation}
where the sum runs over the $i$ energy bins of the $\alpha$ neutrino flavor spectra reported in \cite{IceCube:2020acn,IceCube:2021uhz}, $N_{\text{obs}}(i,\alpha)$ is the estimated experimental total flux in each flavor bin,
\begin{equation}
    N_{\text{obs}}(i,\alpha)=\int_{E_i}\frac{d\Phi_\alpha^{\text{exp}}}{dE_\nu}dE_\nu
\end{equation}
$\sigma^i_{\text{obs}}(i,\alpha)$ is the experimental $1\sigma$ uncertainty on the integrated flux in each flavor bin and $N_{\text{th}}(i,\alpha)$ is the model prediction of the integrated flux for a given set of model parameters
    \begin{equation}
    N_{\text{th}}(i,\alpha)=\int_{E_i}\frac{d\Phi_\alpha(A,\gamma)}{dE_\nu}dE_\nu
\end{equation}
obtained from nuSIprop \cite{Esteban:2021tub}. The muon neutrino data  $N_{\text{obs}}(i,\mu)$ is obtained from the track event data in \cite{IceCube:2021uhz} and the electron and tau neutrino data $N_{\text{obs}}(i,\{e,\tau\})$ is obtained from the cascade event data in \cite{IceCube:2020acn}. For comparison between the model with a resonant $Z^\prime$ boson and the standard neutrino propagation picture, we contrast the $\chi^2(m_{Z'},g',A,\gamma)$ function with the ``Standard Model" value, defined as 
\begin{equation}
   \chi^2_{\text{SM}} =\chi^2(m_{Z'\rightarrow\infty},g'\rightarrow 0,A_0=1.4, \gamma_0=2.6) \,
\end{equation}
where the astrophysical values used correspond to the central values reported in \cite{IceCube:2020acn}. As a crosscheck, we have confirmed that our method reproduces this result using the log-likelihood function as used in \cite{IceCube:2020acn}.
With the IceCube datasets we are using, we find 
\begin{equation}\chi^2_{\text{SM}} = 27.4 
\end{equation}
with 16 data points and 2 model parameters, i.e., 14 degrees of freedom. To compare the improvement of the $Z'$ model to the SM in the astrophysical neutrino data, we make use of the ratio of reduced $\chi^2$ functions, $R_{\chi^2}$ which we define as
\begin{equation}
R_{\chi^2}=\frac{\frac{\chi^2(m_{Z'},g',A,\gamma)}{12}}{\frac{\chi^2_{\text{SM}}}{14}}
\label{eq:ratio}
\end{equation}
which is intended to penalize the introduction of two model parameters with respect to the SM case.% 
\begin{figure}
    \centering
    \includegraphics[scale=0.5]{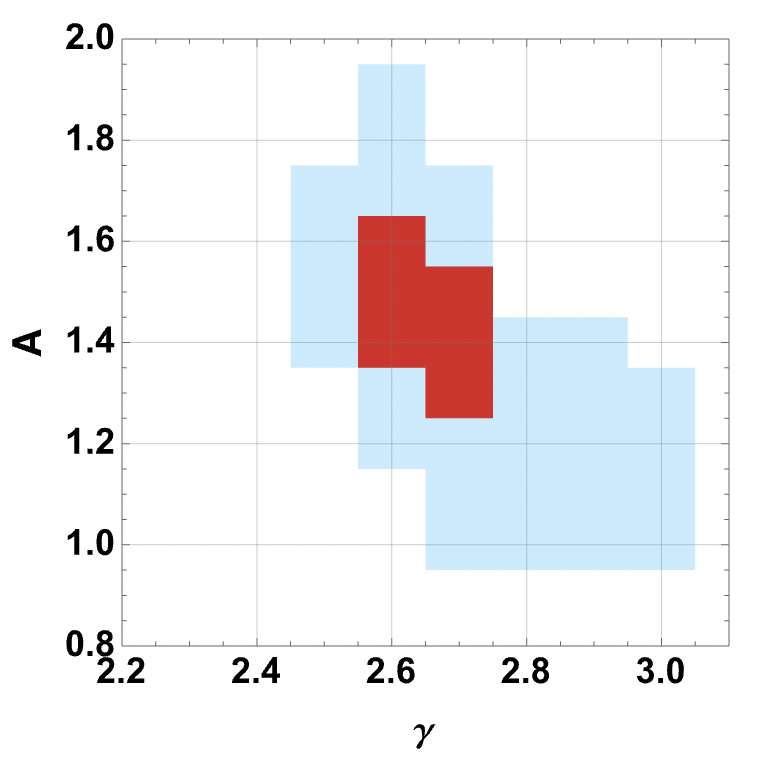}
    \caption{Exclusion plots for astrophysical parameters, marginalized over $m_{Z'}$ and $g'$. The region shaded red (blue) is allowed at 90\% (95\%) C.L.  }
    \label{fig:AvsGamma}
\end{figure}
\section{Results  {\it\&} conclusions}
\begin{figure}
\includegraphics[width=0.5\textwidth]{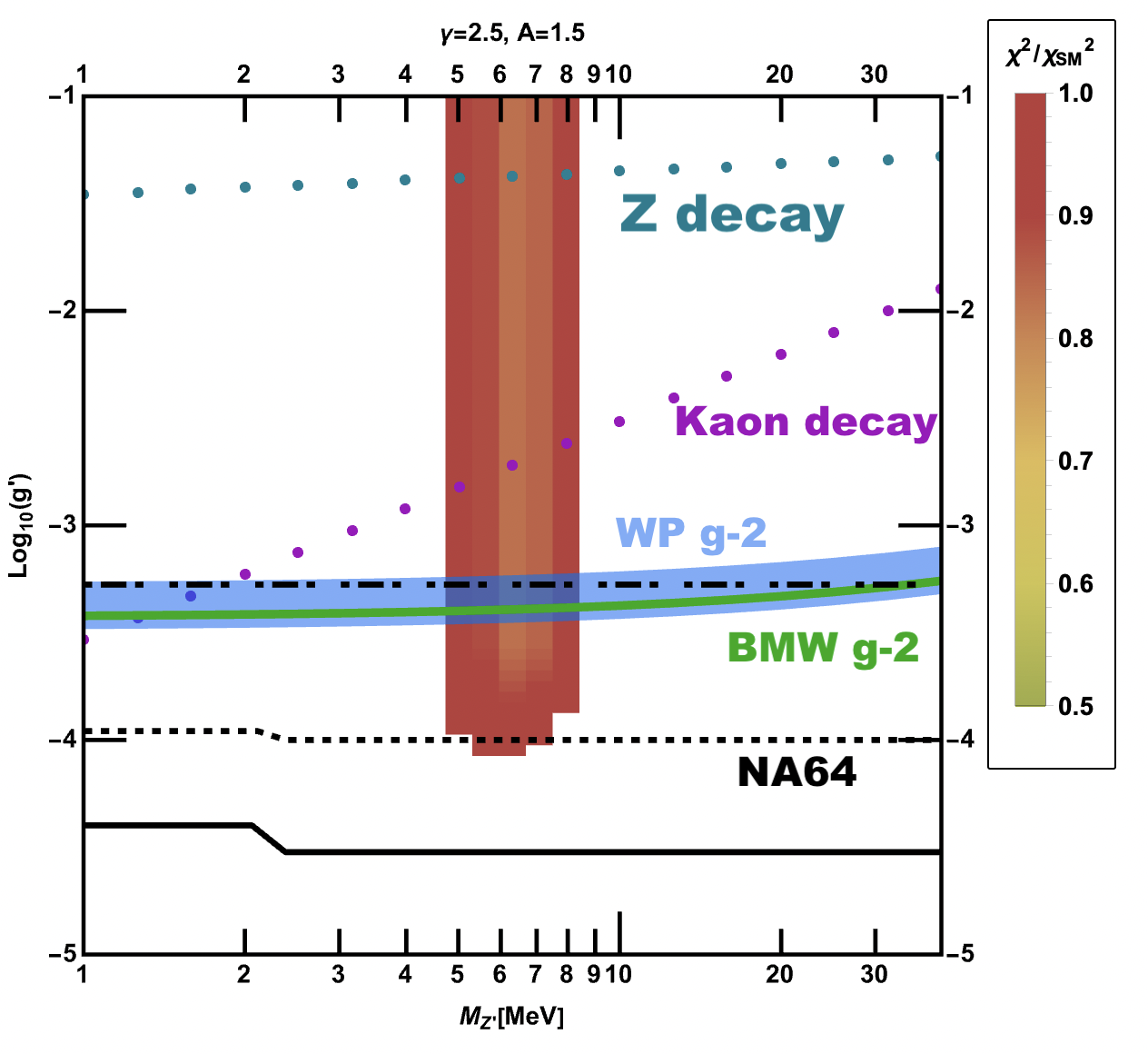}
\caption{Effect of the $Z'$ resonant scattering on the $\chi^2$ function for fixed Astrophysical parameters $(A,\gamma)=(1.5,2.5)$, chosen to contain the global best fit point. The color of the region corresponds to the value of $R_{\chi^2}$. The uncolored region corresponds to values where $R_{\chi^2}$ is larger than 1, equivalently where the effect of the $Z'$ worsens the fit to IceCube data. The region above the blue dotted line is excluded by the limit derived from $Z$ decays.  The region above the purple dotted line is excluded by limits derived from kaon decays. The blue band corresponds to the parameter values that resolve the WP $g-2$ tension while the region above the green line is excluded if the BMW lattice HVP data is contrasted with muon g-2 data.
The black dot-dash-dot line corresponds to the upper limit on $g'$ found by NA64$\mu$ in 2024\cite{Andreev:2024lps}. The black dashed and straight lines correspond to NA64 sensitivity projections for data taking runs with $10^{11}$ and $10^{12}$ muons on target, respectively.  }  
    \label{fig:Mvsg}
\end{figure}

\begin{figure}
    \centering
    \includegraphics[width=0.5 \textwidth]{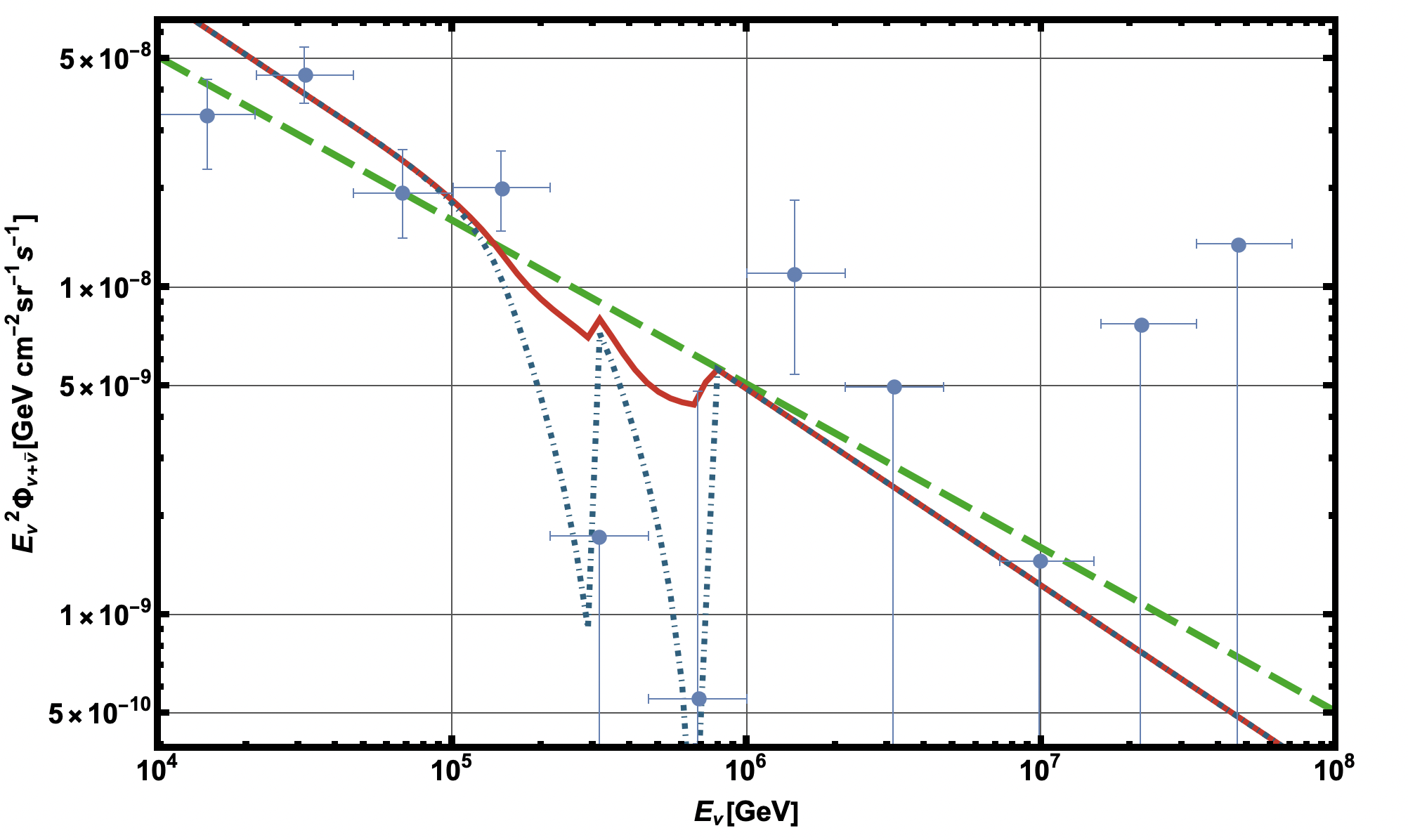}
    \caption{Effect of the $Z'$ resonant scattering on the astrophysical neutrino flux on earth. The dashed green line represents the best fit point obtained by the IceCube collaboration, assuming free neutrino propagation (no $Z'$ interactions), parametrized by $(A,\gamma)=(1.4,2.6)$. The blue dot-dashed and red solid lines correspond to the same astrophysical spectrum, with the effect of resonant $Z^\prime$ interactions.
    The blue dot-dashed line shows the effect of a $Z^\prime$ with a mass $m_{Z^\prime}=6.5$ MeV, and gauge coupling $g^\prime=0.0005$, just at the limit set by NA64 in 2024.  
    The solid line represents instead the spectrum with a mass $m_{Z^\prime}=6.5$ MeV, and gauge coupling $g^\prime=0.0001$, chosen to be at the projected sensitivity of NA64 at $10^{11}$ MOT.  } 
    \label{fig:bfflux}
\end{figure}

Our main results are presented in Figs. \ref{fig:AvsGamma}, \ref{fig:Mvsg} and \ref{fig:bfflux}. In Fig. \ref{fig:AvsGamma} we show contours of the minimum $\chi^2$ function for different fixed values of $(A,\gamma)$ over the range of the parameters $(m_{Z'},g')$ which is allowed by the bounds set by NA64$\mu$, showing the 90\% and 97.5\% Confidence Level (C.L.) contours. Within this region we find two almost degenerate minima. The first minimum, the global best-fit point, corresponds to the parameter set  $P_1 = (m_{Z'},g',A,\gamma)=(6.5\text{ MeV},0.0005,1.5,2.5)$, and the second to $P_2=(m_{Z'},g',A,\gamma)=(6.5\text{ MeV},0.0005,1.4,2.6)$. The values for the astrophysical parameters in these two minima do not depart significantly from the best fit found by the IceCube collaboration analysis \cite{IceCube:2020acn}.

In Fig. \ref{fig:Mvsg} we show the value of the ratio $R_{\chi^2}$, eq. (\ref{eq:ratio}), for fixed values of $(A,\gamma)=(1.4,2.6)$ (containing $P_1$) alongside the limits set by Z decay, Kaon decay, NA64 limits, and projected sensitivities, the band where the $Z'$ solves the muon $g-2$ anomaly, and the exclusion limit derived from muon $g-2$ using the lattice HVP calculation. Inside the uncolored region in this figure, the model predicts a higher reduced $\chi^2$ function than the Standard Model - only hypothesis. It is clear from this figure that if lattice and the new electron-positron data are confirmed, the best-fit point is ruled out. There remains, however some parameter space where the $Z^\prime$ improves the description of IceCube data and its contribution to $(g-2)_\mu$ does not contradict lattice data.  We note that at $10^{11}$ MOT, the NA64 experiment will not only fully probe the $(g-2)_\mu$ region, but also the region where the $Z^\prime$ explains the IceCube neutrino spectrum better than the Standard Model only hypothesis. We note that the range of $M_{Z^\prime}$ masses where the IceCube data is better explained by the resonant scattering is marginally in tension with cosmological constraints on $\Delta N_{\tt eff}$. Increasing the scale of neutrino masses can raise the required $Z^\prime$ mass above 10 MeV, but will then be in tension with cosmological limits on the sum of neutrino masses.

In Fig. \ref{fig:bfflux} we show the neutrino flux on earth obtained for the parameter set ($P_1$), and for the best fit below the future sensitivity of NA64 at $10^{11}$ MOT, alongside the IceCube collaboration best fit to power law spectrum with no $Z^\prime$ interactions. We observe that $P_1$ can indeed fit the observed spectrum better than the Standard Model only hypothesis, and that NA64 at $10^{11}$ MOT has the potential to exclude this $Z^\prime$ model as the source of the IceCube spectrum deficit.

A remark regarding the statistical method used is in order. We found that the $\chi^2$ analysis strategy used here tends to be more sensitive to the anomalous deficit data than the Log-Likelihood analysis done by the IceCube collaboration. This has been observed before in \cite{Hooper:2023fqn,Carpio:2021jhu}. 

With this, we conclude that if the $g-2$ tension disappears and the future results of the NA64 experiments are null, the explanation for the deficit in the UHE neutrino flux must be another mechanism. Given the expected timelines of NA64$\mu$ and IceCube Gen-2, we expect that this scenario will be excluded in the near future by NA64$\mu$, if at $10^{11}$ MOT it fails to detect the signature of the $Z^\prime$. Conversely, if NA64$\mu$ detects a $Z^\prime$ signature, whether IceCube Gen-2 measures a neutrino flux deficit or not will constitute a complementary test of the $Z^\prime$ couplings. 

We have worked in the framework of the $L_\mu-L_\tau$ gauge symmetry, however one could also work in the framework of secret neutrino interactions, where neutrinos acquire $Z'$ couplings through their mixing with heavy neutral leptons charged under the $U(1)'$, thus severely reducing constraints obtained from the $Z'$ coupling to charged leptons while respecting electroweak gauge symmetry.  For example, in the presence of a gauge symmetry that acts on SM singlet fermions, which in turn mix with SM left-handed neutrinos in an inverse seesaw scenario, one can obtain light neutrino couplings to the new gauge boson governed by active-sterile neutrino mixing, while the leading contributions to charged fermion couplings with the new gauge boson arise from gauge boson mixing and one loop effects \cite{Abdallah:2021npg,delaVega:2022uko,Bertuzzo:2018ftf,Bertuzzo:2018itn,Berbig:2020wve,Pasumarti:2023apw}.

Finally, there is also the possibility that the neutrino flux deficit is explained by a multi component source of UHE neutrinos, where the single-power law for the spectrum fails. Several joint analyses of astrophysical events and high-energy neutrino data have been performed with the goal of identifying a single type of event as the main source of the diffuse neutrinos observed in IceCube \cite{Smith:2020oac,M:2024dpj,Lu:2024flp,Bouri:2024ctc,McDonough:2023ngk,Hooper:2023ssc,Chang:2022hqj,Li:2022vsb,Troitsky:2021nvu}. These analyses typically use models of neutrino and photon emission from a proposed type of source, together with point source catalogs and diffuse neutrino and gamma ray or radio data. To date, no singular type of event has been identified as likely responsible for the whole of the diffuse neutrino flux. If multiple types of astrophysical events contribute significantly to the flux, then one might expect that it cannot be described accurately by a single power law. In such a case the resonant effect of the $Z^\prime$ might be difficult to disentangle from features in the initial neutrino spectrum.

\begin{acknowledgments}
This work was supported by the National Science and Technology Council, the Ministry of Education (Higher Education Sprout Project NTU-113L104022-1),  the National Center for Theoretical Sciences of Taiwan,  the University of California Institute for Mexico and the United States (UC MEXUS) (CN 18-128) and the Consejo Nacional de Ciencia y Tecnolog\'{\i}a (CONACYT) (CN 18-128). E.P is grateful for the support of PASPA-DGAPA, UNAM for a sabbatical leave. We thank the referee for useful comments.
\end{acknowledgments}

%\appendix

%\section{Appendixes}
%\subsection{An appendix}
%This is an appendix
%\nocite{*}

\bibliography{apssamp}% Produces the bibliography via BibTeX.

\end{document}